# An initial review of hypersonic vehicle accidents

L. Pollock[1], G. Wild[1]

[1] School of Engineering and Information Technology, UNSW at the Australian Defence Force Academy, Northcott Drive, Campbell, Australian Capital Territory, 2612, Australia

## Abstract

Hypersonic flight, generally defined as the region in which the speed of a vehicle exceeds Mach 5 and in which thermal loads become dominant, has seen more attention over the past several decades due to the potential military applications of such vehicles. Vehicles capable of flight at these speeds, whilst seemingly a novel prospect, have been in development for more than 70 years. The nature of flight in this environment as well as the new challenges it introduces have led to relatively high failure rates. This paper presents a review of hypersonic vehicle flights that have resulted in failure from conception to modern day with the purpose of identifying trends to aid and guide the development of future vehicles. The collected data is used to formulate a failure taxonomy to accurately identify and classify past hypersonic vehicle failures as well as potential future ones. The trends and features of the categorical data collected are explored using an ex-post facto study.

## Introduction

A recent resurgence in the research and development of hypersonic flight vehicles has largely resulted from the announcement of hypersonic weapons development by Russia and China [1]. Whilst the sudden interest in hypersonic flight makes it appear to be novel, it is rather historical and was first achieved in 1949 via the launch of the V-2/WAC Corporal that achieved a top speed of Mach 7.5 [2]. The hypersonic flight regime is broadly defined by many as being at speeds greater than Mach 5; although some indicate that the definition of such flight should not be rigidly applied. As exemplified by Anderson [3], "…hypersonic flow is best defined as that regime where certain physical flow phenomena become progressively more important…" and further, that such flow is not constrained to a terrestrial environment but is vital in the analysis of re-entry vehicles. While hypersonic flight has been studied for over 70 years, very few vehicles have successfully flown in such conditions and even fewer have been reusable. The primary aims of this study is to explore the history of hypersonic vehicle failures in the hope of identifying the prevalent trends and obstacles that oppose such development. Secondarily, the study aims to provide insights to aviation safety organisation with growing civilian space activities (involving re-entry) and potential future hypersonic flight.

## Methodology

To undertake the study, a qualitative ex-post facto approach was adopted [4]. Qualitative data about the failures has largely been obtained from the Astronautix [5] database and supplemented by the NASA historical archives as well as other online sources. Details surrounding specific vehicles as well as the events that resulted in their failure has proved difficult to source, ultimately owing to the sensitive nature of hypersonic technology. Cases where reliable comprehensive records could not be obtained for corroboration by authoritative or multiple independent sources have been excluded. A total of 50 individual failure events have been analysed spanning from 1956 through to 2011. Failures in ground testing have not been included.

**CAST/ICAO ADREP Taxonomy**

Initially the existing taxonomical framework for aircraft was utilised to assess the failure events. Specifically, the Commercial Aviation Safety Team (CAST) and International Civil Aviation Organization (ICAO) Accident/Incident Data Reporting (ADREP) failure occurrence taxonomy was utilised. The ADREP system was developed by the CAST/ICAO Common Taxonomy Team (CICTT) and includes experts from various nations, operations, and occupations [6]. This taxonomy is widely adopted for post-accident analyses and has been utilised for UAS/RPAS [7], maintenance [8], and aircraft structural fatigue failures [9], all technical oriented studies. The ADREP reporting system is a multi-tiered framework that is used to classify the cause of a failure event for further categorical analysis. In the following study, the 50 events were coded as per the ADREP framework. Fig. 1 shown below illustrates a generalised approach to the multi-tier flow for ADREP coding. For example, a failure coded as SCF-NP-AFC-APLS would correspond to autopilot function loss with each coding designation further explained in Fig. 1 and Table 1. The full list of coding designations can be obtained from the CAST/ICAO Common Taxonomy Team website [10, 11]. It is standard practice for multiple codes to apply to single events. A clear example of this is F-POST or fire post impact; clearly something occurred to result in an impact, with the resultant fire causing further effects (damage or injury).

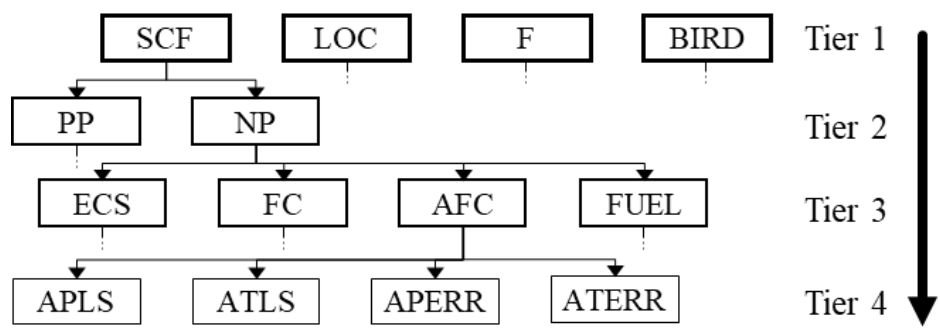

*Fig. 1: Example ADREP taxonomy flow chart coding.*

*Table 1: Example breakdown of the multi-tier flow for ADREP coding.*

| | |
|---|---|
| SCF | System Component Failure |
| NP | Non-Powerplant |
| AFC | Automatic Flight Controls |
| APLS | Autopilot Function Loss |

**Pareto Analysis**

Following coding of the events as per the ADREP framework, a Pareto analysis was performed. The Pareto analysis utilises the Pareto rule, otherwise known as the 80/20 rule owing to the idea that 80% of failures are the result of only 20% of causes [12]. Pareto analyses are a common quality assurance tool that have likewise been widely applied by industry to summarise categorical information graphically [12]. Whilst a useful business tool, Pareto analyses can be applied generally and have previously been used in failure and accident studies [13]. A Pareto chart is similar in appearance to an ordered histogram but likewise includes a cumulative line plot. To ensure the usefulness of the plots the *unofficial* rules of Pareto plots were not followed [14]. A series of Pareto analyses have been conducted upon the ADREP coded failures in the effort of determining the key causes to direct future research and development.

## Results

Fig. 2 below showcases a time-series histogram of the number of hypersonic vehicle failure occurrences. The large peak in failure occurrences during the 1960's is a result of development and testing of the X-15 vehicle that was the focus of hypersonic research at the time. Refocused interest in the space program saw a dramatic drop in pure hypersonics research, as evident by the decrease in events between 1970's and 1990's. The events occurring in the 1980's correspond to Space Shuttle accidents. The renewed interest into hypersonics research becomes more evident with a rise in the 2000's and 2010's; however, to a much lesser degree than previously seen prior to the space program. Hence, whilst there is currently a large focus upon hypersonics research, the lack of failure events indicates that few fully integrated and reusable vehicles have been developed.

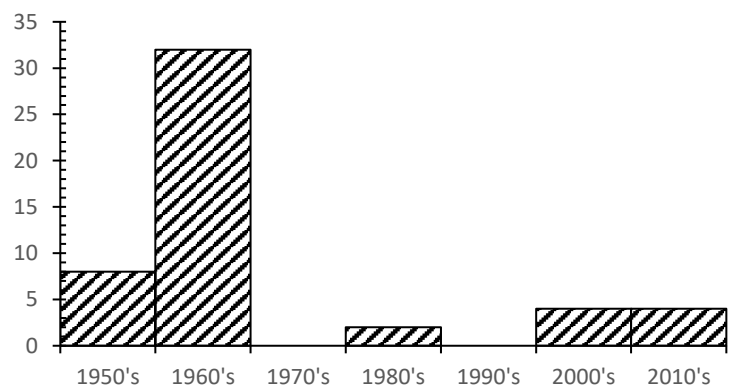

*Fig. 2: Example ADREP taxonomy flow chart coding.*

Fig. 3 illustrates a Pareto plot of generalised hypersonic failure coding results dependent upon the specific events that proceeded the failure. Fig. 4 presents the Pareto plot of the ADREP coded failure occurrences. The 80% line indicates that specific focus should be addressed towards system component failures, both non-power plant systems and power plant systems.

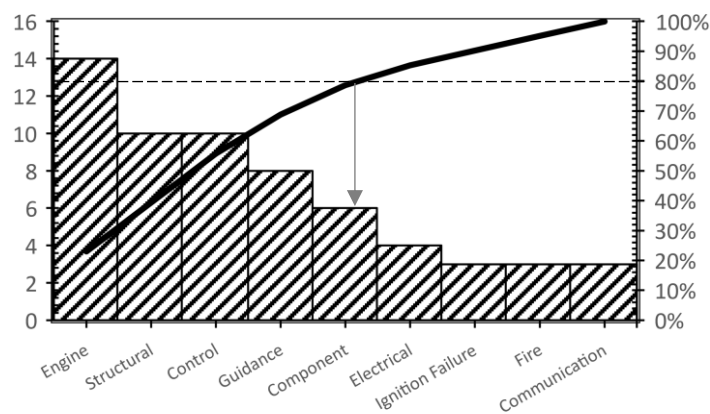

*Fig. 3: A Pareto plot of broad hypersonic occurrence categories.*

A further analysis of both powerplant and non-powerplant failures is shown in Fig. 5 below. Application of the 80% rule once again, indicates that for non-powerplant failures, structural, avionic, and landing gear failures are most prevalent. Further investigation indicates that all landing gear failures occurred within X-15 flights and that all avionic failures have been attributed to navigational information loss due to guidance issues. Five of the six LG failures were specifically uncommanded gear deployment. The taxonomy specifically states to code these as "other", hence the taxonomy is not lacking here. For powerplant failures, both in-flight shutdown events and loss of engine control require particular focus.

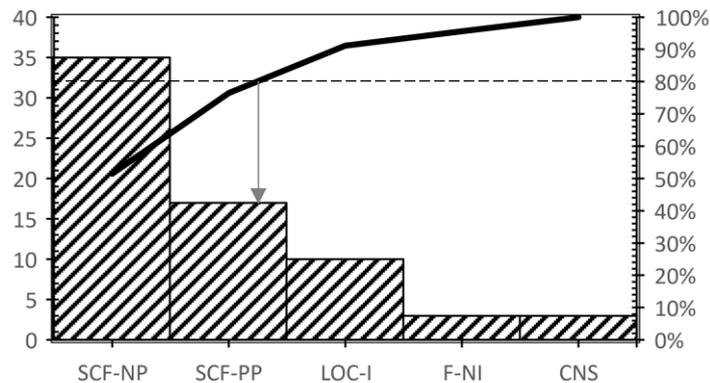

*Fig. 4: A combined tier one and two Pareto plot of ADREP coded failures (Tier 1).*

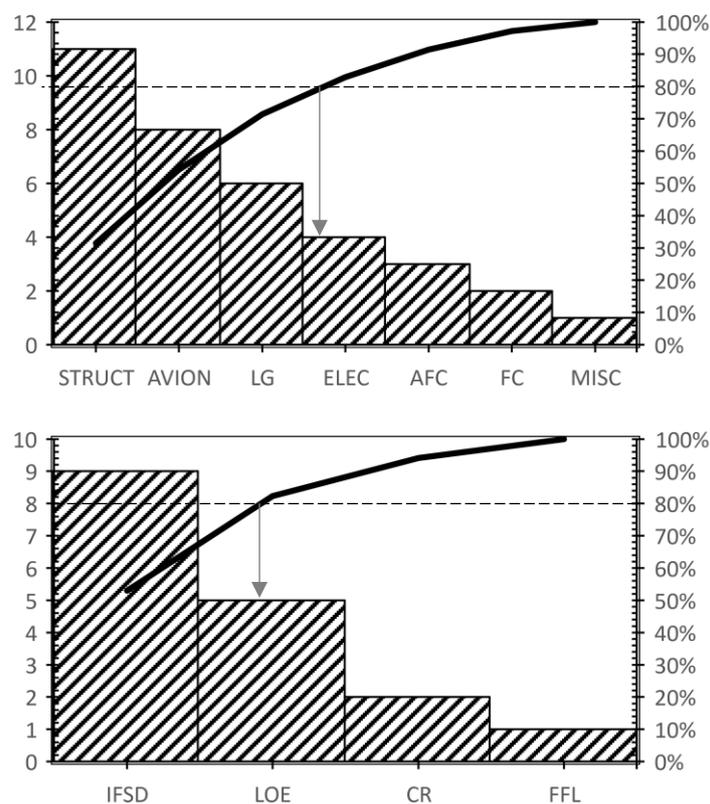

*Fig. 5: Pareto plot breakdowns for non-powerplant (top) and powerplant (bottom) failures (Tier 3).*

Results from the Pareto analysis of the structural occurrence failures, as shown in Fig. 6, highlight that the dominant causes include control surface failure, cracking, cabin window damage, and "other" events. The "other" event category, being of highest occurrence, is a prime indicator of the need to develop unique taxonomical classification for hypersonic failure events that otherwise do not conform to the current ADREP system. The final group of Tier 4 data is two electrical categories, SCF-NP-ELEC-GENLS (three cases) and SCF-NP-ELEC-SYSMALF (one case). Given many electrical and avionics systems are similar between conventional and hypersonic aircraft, there was no spike in the "other" categories for these types of failures.

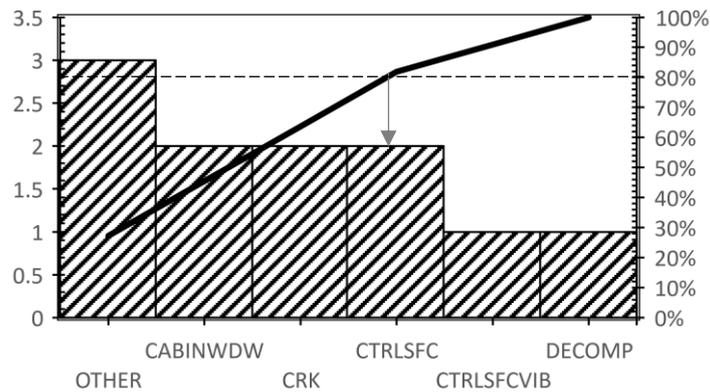

*Fig. 6: A Pareto plot of the structural occurrence failures (Tier 4).*

## Discussion

The application of the ADREP taxonomy results in the loss of hypersonic specific failure modes. This clearly indicates that improvements should be included to account for issues specific to reusable launch and hypersonic vehicles. This is essential given the recent rapid developments is modern space tourism from suborbital [15], in July of 2021 to orbital [16], in September of 2021. Firstly, due to the use of rocket propulsion systems, a Tier 3 code for ignition failure should be added, potentially with Tier 4 codes indicating the type of ignition failure. Interestingly ignition systems are an essential engine system [17], and even air transport pilots learn about them, yet there is no code in the ADREP taxonomy for the failure of any component of the ignition system, even though in reciprocating engines spark failures are common [18, 19]. Historical evidence indicates that Tier 4 codes would primarily constitute premature or failed ignition; however, more may be necessary. The proposed code for this failure occurrence is in the form of SCF-PP-IGNI-XX.

Secondly, a Tier 3 code for explosive failure should be included. The occurrence of explosive failures would aid in categorising those events that cannot be traced to a primary failure point owing to the extensive damage caused by the explosion. Further Tier 4 coding could be applied should more information be known regarding the cause of the explosion, e.g., fuel leak, but of which the cause of the fuel leak is not known. The proposed code for this failure occurrence is in the form of SCF-PP-EXP-XX. Related to this is the third recommendation, for the addition of coding that would identify at which stage of the vehicle the failure occurred. This would allow for classification and greater analysis of multiple stage vehicles that employ several propulsion systems. The explosion of the final stage hypersonic vehicle has radically different consequences than the explosion of the first stage launch vehicle.

Finally, owing to the large number of STRUCT-OTHER failure events that were difficult to specifically classify, two Tier 4 codes are proposed that would allow for the majority attribution of these events. The proposed Tier 4 codes are that of TPS which would constitute damage to the vehicle's thermal protection system and that of THERM, which would constitute degradation or failure of the vehicle structure owing to excessive aerodynamic heating. Examples of THERM events may include those such as excessive heating that results in substantial and unexpected mechanical property reductions or those that result in ablation or corrosion to the structure. While both a TPS and a THERM event could result from aerodynamic heating, a TPS event would be the failure of part of the vehicle where failure was not expected due to the presence of the protection system. For a THERM event, the failure would be due to unexpected heating where the TPS is not present.

# Conclusion

This study has employed an ex-post facto analysis of hypersonic vehicle flight failures through the application of Pareto analysis. The analysis has indicated that hypersonic vehicles demonstrate poor powerplant and non-powerplant system component reliability. The assessment of non-powerplant failures indicates that specific attention should be provided towards improving structural and guidance reliability. For powerplant failures, specific interest should be provided to preventing loss of in-flight shutdown events and loss of engine control. The study has likewise revealed that the ADREP taxonomy is inadequately designed for the safety assessment of hypersonic vehicles and several recommendations have been made to further develop the classification system. Due to the often-sensitive nature of hypersonic vehicle research, there have been large limitations in accessing comprehensive failure records. Future work is planned by the authors to expand the scope of this study to include launch vehicles and to provide greater quantitative analysis.